\documentclass[twocolumn,english,prx,notitlepage,superscriptaddress,longbibliography]{revtex4-2}
\usepackage{graphicx}
\usepackage{amsmath}
\usepackage{amssymb}
\usepackage{bm}
\usepackage{enumerate}
\usepackage{braket}
\usepackage{xfrac}
\usepackage[caption=false]{subfig}
\usepackage[normalem]{ulem}
\usepackage[usenames,dvipsnames]{xcolor}
\usepackage{txfonts}
\usepackage[mathscr]{euscript}
\usepackage[colorlinks = true,
linkcolor = blue,
urlcolor  = blue,
citecolor = blue,
anchorcolor = blue]{hyperref}
\usepackage{verbatim}
\usepackage{ulem}

\begin{document}

\title{Spectrum and low-temperature bulk properties of triangular quantum spin liquid candidate NaYbSe$_2$}

\author{A. O. Scheie}
\email{scheie@lanl.gov}
\affiliation{Los Alamos National Laboratory, Los Alamos, NM 87545, USA}

\author{Minseong Lee} 
\email{ml10k@lanl.gov}
\affiliation{National High Magnetic Field Laboratory, Los Alamos National Laboratory, Los Alamos, New Mexico 87545, USA}

\author{Kevin Wang} 
\affiliation{Department of Physics, University of California, Berkeley, CA 94720, USA}

\author{P. Laurell}
\affiliation{Department of Physics and Astronomy, University of Tennessee, Knoxville, TN 37996, USA}

\author{E. S. Choi} 
\affiliation{National High Magnetic Field Laboratory, Florida State University, Tallahassee, FL 32310, USA}

\author{D. Pajerowski} 
\affiliation{Neutron Scattering Division, Oak Ridge National Laboratory, Oak Ridge, TN 37831, USA}

\author{Qingming Zhang} 
\affiliation{School of Physical Science and Technology, Lanzhou University, Institute of Physics, Chinese Academy of Sciences, Lanzhou 730000, China}

\author{Jie Ma} 
\affiliation{Department of Physics and Astronomy, Shanghai Jiao Tong University, Shanghai 200240, China}

\author{H. D. Zhou} 
\affiliation{Department of Physics and Astronomy, University of Tennessee, Knoxville, TN 37996, USA}

\author{Sangyun Lee}
\affiliation{National High Magnetic Field Laboratory, Los Alamos National Laboratory, Los Alamos, New Mexico 87545, USA}
\affiliation{National High Magnetic Field Laboratory High B/T Facility, University of Florida, Gainesville, FL, USA}
\affiliation{Department of Physics, University of Florida, Gainesville, FL, USA}
\author{Chao Huan}
\affiliation{National High Magnetic Field Laboratory High B/T Facility, University of Florida, Gainesville, FL, USA}
\affiliation{Department of Physics, University of Florida, Gainesville, FL, USA}

\author{S. M. Thomas}
\affiliation{Los Alamos National Laboratory, Los Alamos, NM 87545, USA}
\author{M. O. Ajeesh}
\affiliation{Los Alamos National Laboratory, Los Alamos, NM 87545, USA}
\author{P. F. S. Rosa}
\affiliation{Los Alamos National Laboratory, Los Alamos, NM 87545, USA}

\author{Ao Chen}
\affiliation{Theoretical Physics III, Center for Electronic Correlations and Magnetism, Institute of Physics, University of Augsburg, D-86135 Augsburg, Germany}

\author{Vivien S. Zapf} 
\affiliation{National High Magnetic Field Laboratory, Los Alamos National Laboratory, Los Alamos, New Mexico 87545, USA}

\author{M. Heyl}
\affiliation{Theoretical Physics III, Center for Electronic Correlations and Magnetism, Institute of Physics, University of Augsburg, D-86135 Augsburg, Germany}

\author{C. D. Batista}
\affiliation{Department of Physics and Astronomy, University of Tennessee, Knoxville, TN 37996, USA}
\affiliation{Neutron Scattering Division, Oak Ridge National Laboratory, Oak Ridge, TN 37831, USA}

\author{E. Dagotto}
\affiliation{Department of Physics and Astronomy, University of Tennessee, Knoxville, TN 37996, USA}
\affiliation{Materials Science and Technology Division, Oak Ridge National Laboratory, Oak Ridge, TN 37831, USA}

\author{J. E. Moore}
\email{jemoore@berkeley.edu}
\affiliation{Department of Physics, University of California, Berkeley, CA 94720, USA}

\author{D. Alan Tennant}
\email{dtennant@utk.edu}
\affiliation{Department of Physics and Astronomy, University of Tennessee, Knoxville, TN 37996, USA}
\affiliation{Department of Materials Science and Engineering, University of Tennessee, Knoxville, TN 37996, USA}

\date{\today}


\begin{abstract}
  We report neutron scattering, pressure-dependent AC calorimetry, and AC magnetic susceptibility measurements of triangular lattice NaYbSe$_2$. We observe a continuum of scattering, which is reproduced by matrix product simulations, and no phase transition is detected in any bulk measurements. Comparison to heat capacity simulations suggest the material is within the Heisenberg spin liquid phase. AC Susceptibility shows a frequency-dependent peak at 40~mK, as has been observed in several triangular magnets. 
\end{abstract}
\maketitle


A quantum spin liquid (QSL) is a state of matter first predicted by P. W. Anderson in 1973, wherein spins arranged in a lattice exhibit a massively entangled and fluctuating ground state \cite{Anderson1973}. One defining characteristic of QSLs is their fractional excitations, which interact with each other through emergent gauge fields \cite{Savary_2016review,broholm2019quantum}. The potential for topological protection from decoherence makes QSLs appealing platforms for quantum technologies. However, despite decades of searching and extensive theoretical work, no unambiguous examples of a quantum spin liquid material have been found.

Anderson's original prediction for a QSL state was the two-dimensional triangular lattice antiferromagnet. With nearest-neighbor exchange only, this system orders magnetically, but a small antiferromagnetic second-nearest-neighbor exchange $J_2$ theoretically stabilizes a QSL phase \cite{PhysRevB.92.041105,PhysRevB.92.140403,PhysRevB.93.144411,PhysRevB.94.121111,PhysRevB.95.035141,PhysRevB.96.075116,PhysRevLett.123.207203}. 
Though the existence of this phase is well-accepted theoretically, it is not clear 
what kind of QSL such a state would be. Proposals include a gapless $\mathbb{U}_1$ Dirac QSL \cite{PhysRevB.93.144411,PhysRevLett.123.207203,doi:10.7566/JPSJ.83.093707,drescher_dynamical_2023}, 
a valence bond crystal \cite{PhysRevX.14.021010,seifert2023spin},  
a gapped $\mathbb{Z}_2$ QSL \cite{Sachdev92,PhysRevB.92.041105,PhysRevB.92.140403,PhysRevB.107.L140411}, 
or a chiral spin liquid \cite{PhysRevLett.127.087201}.  
Because numerical simulations are limited by finite size, theoretical results are ambiguous \cite{Sherman_2023_spectral,PhysRevX.14.021010}. The best (and perhaps only) way to resolve this question would be to find a real material which harbors the triangular lattice QSL ground state. 

Inelastic neutron scattering studies of triangular antiferromagnets with nearest-neighbor exchange have revealed anomalous continuum scattering that cannot be explained by semiclassical theories~\cite{Ma16,Ito17,Macdougal20}. The measured single-magnon dispersion was accurately reproduced using a Schwinger Boson approach, where magnons are obtained as two-spinon bound states \cite{Ghioldi_2018,Ghioldi22}. This suggests that these ordered magnets are in close proximity to a gapped $\mathbb{Z}_2$ QSL as the deconfined Schwinger Boson phase. But an unambiguous measurement of a material in the deconfined QSL phase has not been reported. 

\begin{figure}
	\centering
	\includegraphics[width=0.47\textwidth]{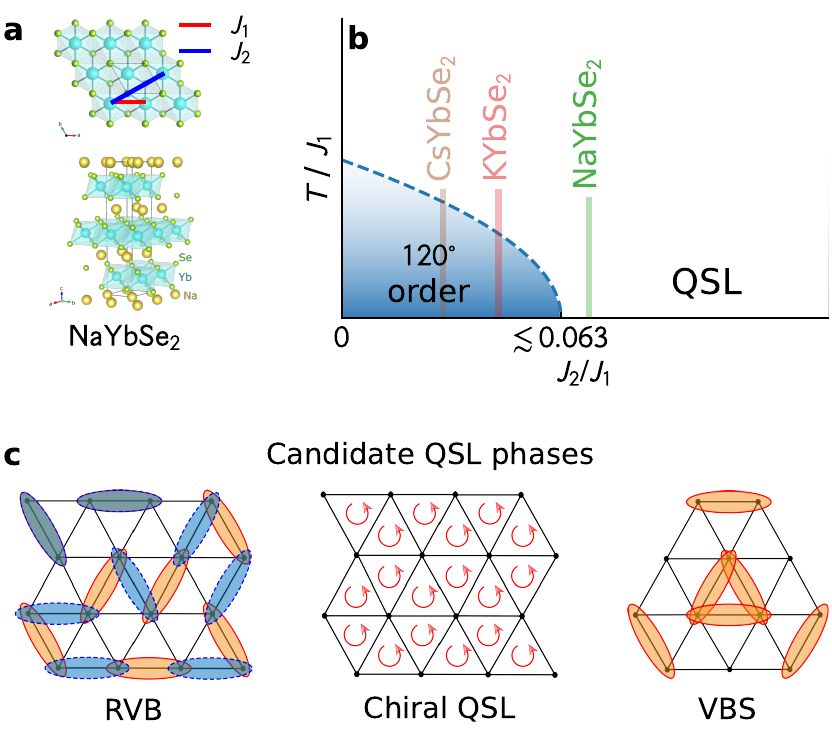}
	\caption{Triangular lattice quantum spin liquid (QSL). Panel \textbf{a} shows the  NaYbSe$_2$ crystal structure. Panel \textbf{b} shows the conceptual phase diagram of CsYbSe$_2$, KYbSe$_2$, and NaYbSe$_2$ as a function of fitted $J_2/J_1$ values \cite{Scheie_2024_Nonlinear}. The theoretical boundary to the quantum spin liquid phase for the isotropic model is detected by neural quantum state (NQS) simulations at $J_2/J_1 \lesssim 0.063 \pm 0.001$ (see Supplemental Materials) locating NaYbSe$_2$ well within the QSL phase.  
    Panel \textbf{c} shows schematics for potential gapped phases on the triangular lattice: the gapped $\mathbb{Z}_2$ QSL as resonating valence bond (left), a chiral QSL (center), and a particular ordering pattern \cite{seifert2023spin} for a 12-site valence bond solid (right). Note: overlapping ovals represent resonating singlet bonds.
    }
	\label{fig:Schematic}
\end{figure}

A very promising class of materials is the Yb delafossites $A$YbSe$_2$ where $A$ is an alkali metal \cite{Ranjinth2019,Ranjith2019_2,Zhang_2021_NYS,Dai_2021,scheie2024_KYS,Xie2023}. These form ideal triangular lattices of magnetic Yb$^{3+}$, and appear to approximate the Heisenberg $J_2/J_1$ model \cite{Scheie_2024_Nonlinear}, see Fig. \ref{fig:Schematic}. 
Of these, CsYbSe$_2$ and KYbSe$_2$ have been observed to order magnetically at zero field \cite{scheie2024_KYS,Xie2023}. However, following a trend in the periodic table that a smaller $A$-site element enhances $J_2$ and destabilizes order \cite{Scheie_2024_Nonlinear}, no long-range magnetic order has been observed in NaYbSe$_2$ \cite{Ranjinth2019,Dai_2021}, which makes it a prime candidate for a QSL ground state. 
Importantly, the tunability of these compounds means that the QSL phase can be approached systematically from magnetic order (Fig. \ref{fig:Schematic}).  This allows for greater confidence and rigor than studying a single compound in isolation would. 

Previous NaYbSe$_2$ studies reported a diffuse neutron spectrum that was interpreted in terms of spinon Fermi surface excitations from a QSL \cite{Dai_2021}, but because of 3\% Na site disorder on those samples, it is not clear whether the magnetic order and coherent excitations were destroyed by small amounts of disorder (as in the ill-fated Yb$^{3+}$ QSL candidates YbMgGaO$_4$ \cite{Paddison2017} and Yb$_2$Ti$_2$O$_7$ \cite{Ross_2011_YTO}). 
To further clarify NaYbSe$_2$, we measured the inelastic neutron spectra, AC calorimetry, and AC susceptibility with high quality samples. We observe coherent excitations, lack of magnetic order, and a 40~mK frequency-dependent freezing transition. This is strong evidence for a QSL ground state in NaYbSe$_2$ that is disrupted by dilute randomness or impurities at the lowest energies. 



\begin{figure*}
	\centering
	\includegraphics[width=\textwidth]{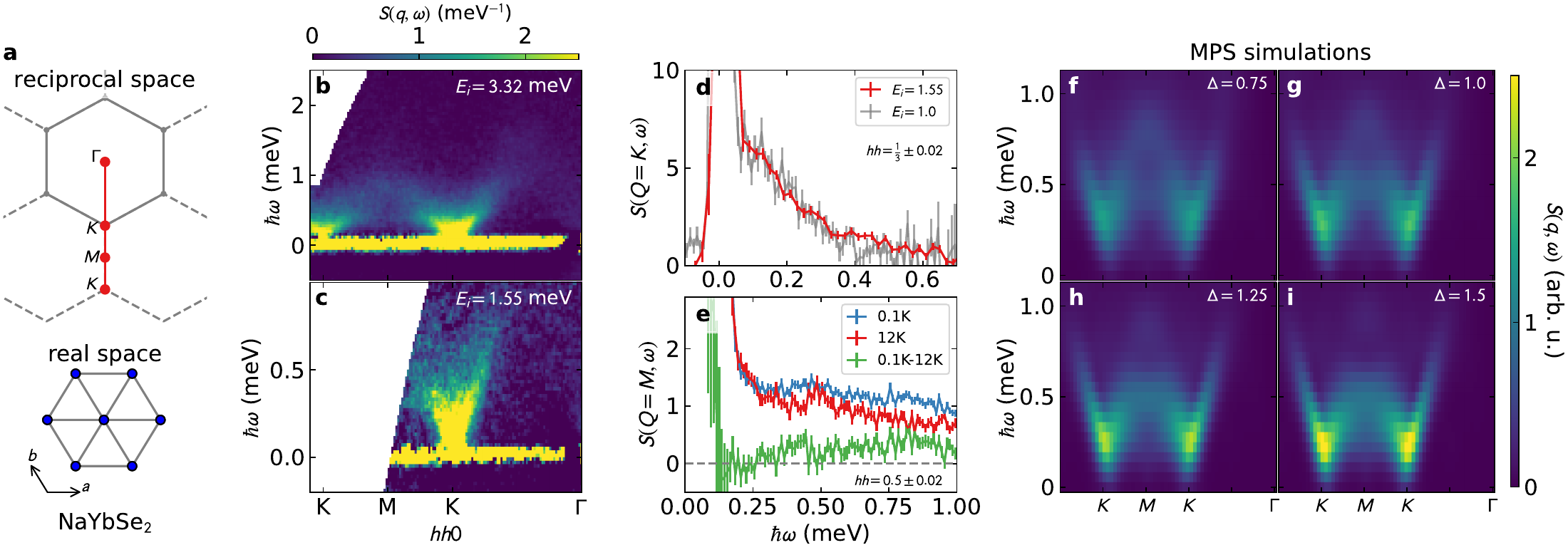}
	\caption{Neutron scattering data on NaYbSe$_2$. Panel \textbf{a} shows the triangular crystal structure and reciprocal space vectors, with $hh$ vertical. Panels \textbf{b} and \textbf{c} show neutron scattering in the $hh\ell$ scattering plane, integrated over $\ell < 4.5$ reciprocal lattice units (rlu) with incident neutron energies $E_i = 3.32$~meV and $1.55$~meV respectively. 12~K data have been subtracted as a background, see supplemental materials.
    Panel \textbf{d} shows the scattering at K$=(1/3,1/3,0)$ as a function of energy with $E_i = 1.0$~meV and $1.55$~meV. To an energy resolution of 50~$\rm \mu$eV, the spectrum is gapless. 
    Panel \textbf{e} shows the temperature-subtracted scattering at M$=(1/2,1/2,0)$ with $E_i=3.32$~meV, and it is unclear whether the spectrum is gapped or gapless at M. 
    Panels \textbf{f}-\textbf{i} show MPS simulated scattering of NaYbSe$_2$ with varying levels of anisotropy. Note the broadened signal due to finite size effects. 
    }
	\label{fig:NeutronSubtractionMain}
\end{figure*}

The neutron spectra at 100~mK, shown in Fig. \ref{fig:NeutronSubtractionMain}, show a highly dispersive continuum of excitations with a well-defined lower bound, similar to KYbSe$_2$ \cite{scheie2024_KYS} (see supplemental materials for experimental details). This is qualitatively different from the spectra measured by Dai et al \cite{Dai_2021} on the 3\% Yb/Na site-mixed sample which in contrast showed smeared our continua in k-space and diffuse spectra extending to low energies in many regions of reciprocal space. (Later in the text, we will explain why we believe our samples have lower levels of mixing disorder.)
Here, the only region of reciprocal space which has appreciable intensity down to low energies is $(1/3,1/3,0)$, corresponding to the 120$^\circ$ magnetic order seen in sister compounds KYbSe$_2$ \cite{scheie2024_KYS} and CsYbSe$_2$ \cite{Xie2023}. Down to 50~${\rm \mu eV}$ (the limit before the incoherent scattering on the elastic line obscures the scattering energy for the incident energy of $E_i=1$~meV), no gap in the spectrum is resolved. 

For comparison we also show matrix product state (MPS) calculated spectra in \ref{fig:NeutronSubtractionMain}\textbf{f}-\textbf{i} with $J_2/J_1 = 0.071$ (this value derived from finite field non-linear spin wave fits \cite{Scheie_2024_Nonlinear}), at varying levels of exchange anisotropy $\Delta$ (see Supplemental Materials). The boundary to the quantum spin liquid phase for the isotropic model is at $J_2/J_1=0.063$ calculated using neural quantum states (see Supplemental Materials) locating the material in the theoretically predicted QSL phase for weak anisotropies.  
Because of finite size lattice effects the calculated spectra are gapped, and it is difficult to make quantitative comparisons between theory and neutron experiments. Nevertheless, the calculated spectra are consistent with the observed spectra, corroborating the idea that a $J_2/J_1$ model with easy-plane anisotropy is an appropriate model for NaYbSe$_2$.

\begin{figure*}
	\centering
	\includegraphics[width=0.96\textwidth]{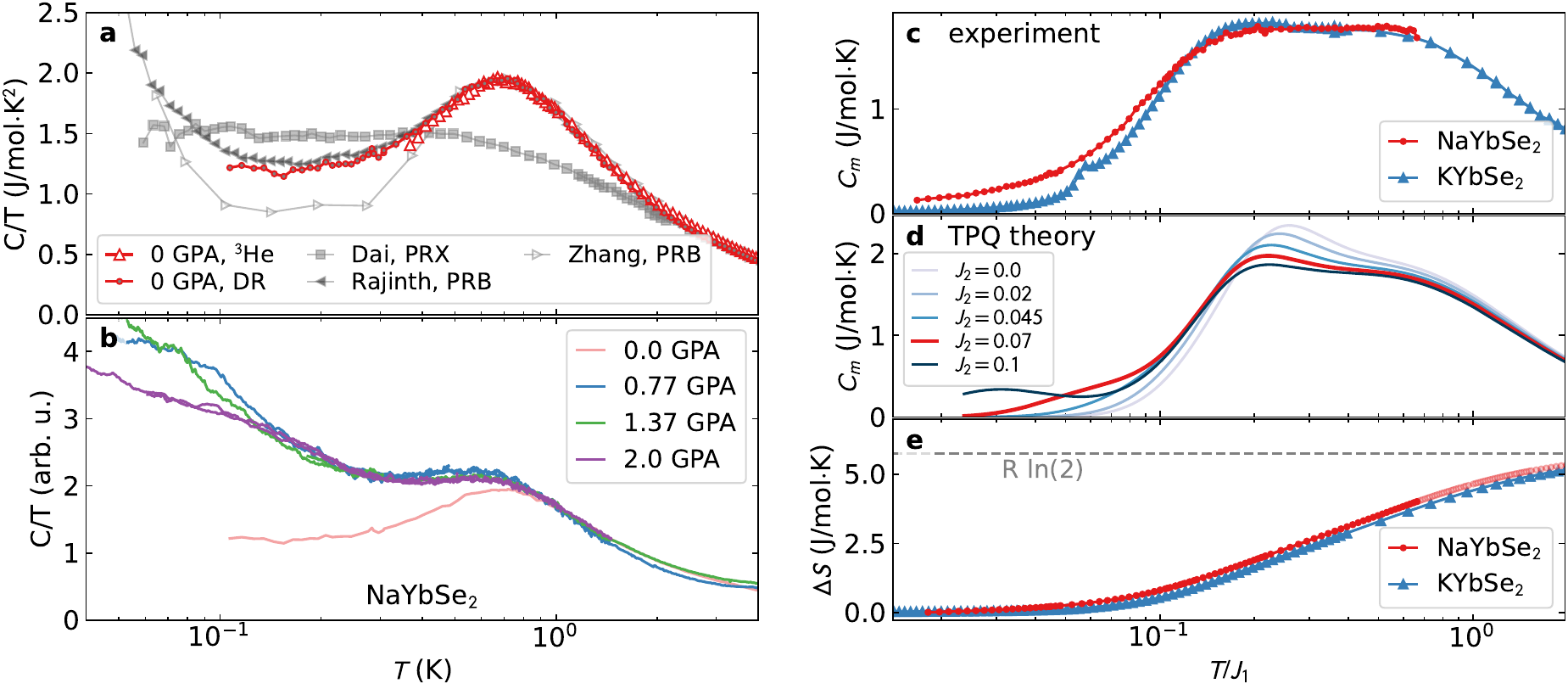}
	\caption{ NaYbSe$_2$ specific heat. Panel \textbf{a} shows ambient pressure specific heat, measured with both a $^3$He PPMS insert and the dilution refrigerator (DR). In grey are the data from previous studies \cite{Dai_2021,Ranjith2019_2,Zhang_2021_NYS-HC}. Panel \textbf{b} shows pressure dependent heat capacity. The low-temperature upturn is an artifact of 
 measuring in a pressure medium with finite thermal conductivity, 
 but no pressure-induced magnetic ordering transition is visible in the data. Panel \textbf{c} shows the magnetic specific heat of NaYbSe$_2$ compared to KYbSe$_2$ \cite{scheie2024_KYS}, with the temperature axis scaled by $k_B J_1$ for each compound. Panel \textbf{d} shows the theoretical calculated specific heat from TPQ (see text) as a function of $J_2$ in units of $J_1$. The theoretical trend confirms that NaYbSe$_2$ is closer to the QSL. Panel \textbf{e} shows the integrated entropy, revealing that both compounds converge close to $R\ln(2)$.}
	\label{fig:HC}
\end{figure*}

Despite intensity concentrated at $(1/3,1/3,0)$ and similar spectra to CsYbSe$_2$ and KYbSe$_2$, we observe no static magnetic order in NaYbSe$_2$ in neutron scattering measurements down to 100~mK. 
No magnetic ordering features are visible in heat capacity down to 100~mK either, as shown in Fig. \ref{fig:HC}. (Note also that our sample has similar low-temperature specific heat to those reported in Refs. \cite{Ranjith2019_2,Zhang_2021_NYS-HC}. If the $C/T$ maximum at 800~mK is an indication of sample quality, our sample is free from the site mixing reported in Ref. \cite{Dai_2021}.)
To test whether applied hydrostatic pressure can induce order---as in KYbSe$_2$ wherein pressure enhanced $T_N$ \cite{Scheie_2024_Nonlinear}---we also measured AC calorimetry under pressure (see Supplemental Materials) shown in Fig. \ref{fig:HC}\textbf{b}. Up to 2.0~GPa, no sharp feature as expected for an ordering transition is seen in the data (pressure-dependent thermalization issues cause the low-$T$ specific heat to increase at low $T$, but this is a known artifact and would not mask a sharp ordering transition).

Also in  Fig. \ref{fig:HC}\textbf{c} we compare NaYbSe$_2$ heat capacity to KYbSe$_2$, with the temperature axis rescaled by the fitted $J_1$  \cite{Scheie_2024_Nonlinear}. This shows not only a lack of ordering transition, but also a smaller $k_B T/J_1 \approx 0.2$ maximum heat capacity and greater low-temperature heat capacity in  NaYbSe$_2$ relative to  KYbSe$_2$. 
Comparing this to  thermal pure quantum state (TPQ) simulations of the 27-site 2D triangular lattice in Fig. \ref{fig:HC}\textbf{d}, these trends are beautifully explained with a larger $J_2/J_1$ in  NaYbSe$_2$: the low-temperature heat capacity is largest when $J_2/J_1 \approx 0.07$ and the $k_B T/J_1=0.2$ bump is suppressed with larger $J_2$. Because the TPQ simulations are of a finite size cluster which induces an artificial energy gap, the lowest temperature trends are not quantitatively accurate. However, on a qualitative level, this is remarkable confirmation that NaYbSe$_2$ is indeed closer to or inside the triangular QSL phase.

To investigate the magnetic state to lower temperatures, we measured AC susceptibility down to 20~mK with AC and DC field applied along the $a$ and $c$ directions on NaYbSe$_2$, and again down to 0.7~mK in zero-field for the AC field along $a$ (see Supplemental Materials). 
In this case we observe a clear magnetization plateau in the $B \parallel a$ direction at 5~T, but not for $B \parallel c$ (note these data were collected simultaneously on two separate crystals mounted on two separate susceptometers mounted on the same dilution refrigerator). This agrees with previous measurements \cite{Ranjith2019_2}, and indicates an easy-plane exchange anisotropy in NaYbSe$_2$: in the perfectly isotropic triangular model, 1/3 magnetization plateaux appear both in-plane and out-of-plane, but the out-of-plane plateau is suppressed by planar anisotropy \cite{Chubukov_1991,Sellmann_2015,Quirion_2015}, although the in-plane magnetism still has a continuous rotation symmetry and similar physics is preserved.

\begin{figure*}
	\centering
	\includegraphics[width=\textwidth]{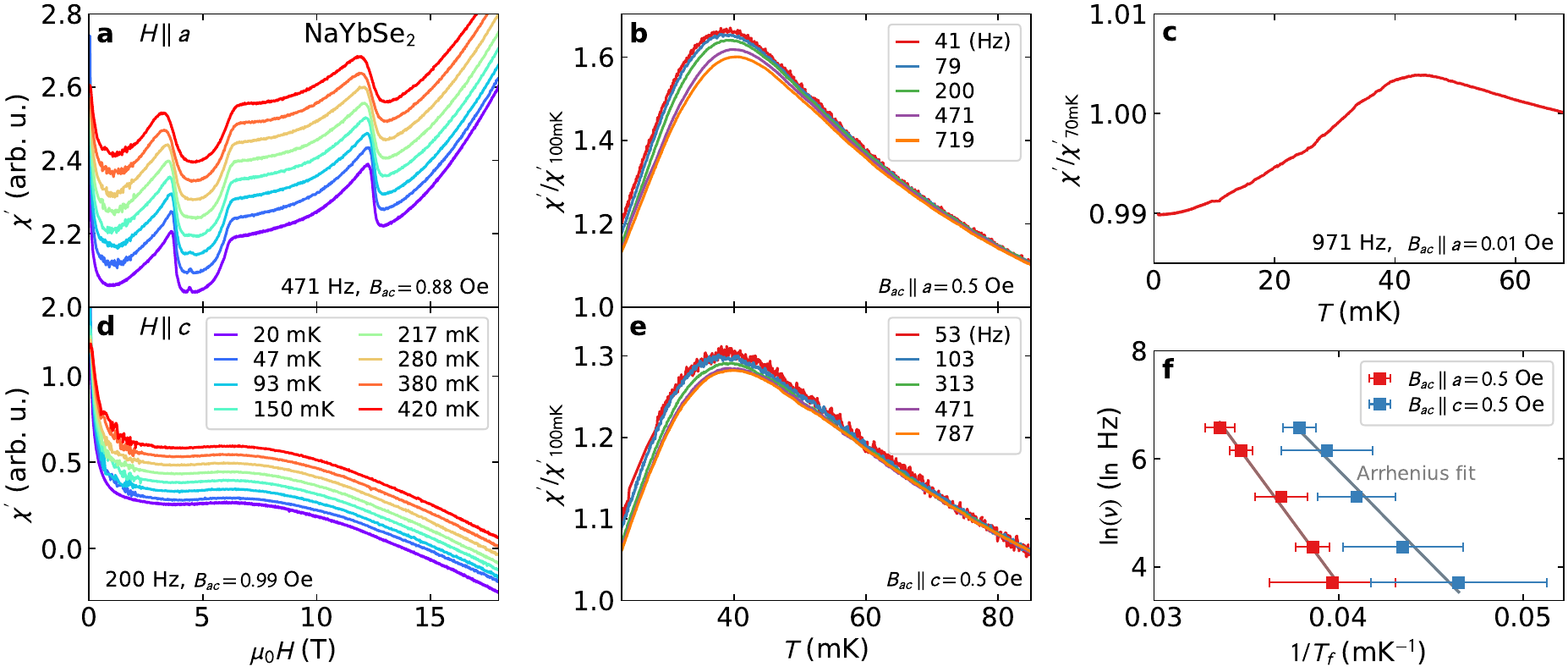}
	\caption{Magnetic susceptibility of NaYbSe$_2$ with $B \parallel a$ (\textbf{a}-\textbf{c}) and $B \parallel c$ (\textbf{d}-\textbf{e}). The left panels (a and d) show the field-dependence, showing a 5~T drop in susceptibility from a magnetization plateau in-plane, but no plateau out-of-plane. The center panels (\textbf{b} and \textbf{e}) show the temperature dependence at low fields and various frequencies, with a symmetric frequency-dependent peak at 40~mK. Panel \textbf{c} shows temperature dependent susceptibility measured in zero dc magnetic field down to 0.7~mK with an excitation field of 0.01~Oe, showing a broad maximum around 40 mK. Panel \textbf{f} shows Arrhenius analysis of the characteristic temperature $T_f$ defined by the peak position of the in-plane and the out-of-phase susceptibility $\chi^{\prime\prime}$. The data follow an Arrhenius behavior, $\ln(\nu) \propto 1/T_f$.}
	\label{fig:susceptibility}
\end{figure*}

Susceptibility also shows a low-field peak at 40~mK as shown in Fig. \ref{fig:susceptibility}\textbf{b} and \textbf{e}. 
(Originally this was misinterpreted to be a gap when measuring with a 0.9~Oe excitation field see the Supplemental Information.)  
The frequency-dependence in-plane and out-of-plane can be quantified by defining a characteristic temperature $T_f$ at the peak position of $\chi^{\prime\prime}$, which follows an activated form described by the Arrhenius relation
\begin{equation}
    \nu=\nu_0\exp \left(-\frac{E_a}{k_B T_f}\right),
\end{equation}
where $\nu_0$ is the attempt frequency and $E_a$ is an effective activation energy. For $B_{ac}\parallel a$, the data are well described by $\nu_0\approx3\times{10}^8$~Hz and $E_a/k_B\approx345$~mK, while for $B_{ac}\parallel c$ we obtain $\nu_0\approx3.6\times{10}^8$~Hz and a slightly larger activation scale $E_a/k_B\approx463$~mK. Moreover, fitting the data to a Vogel–Fulcher form, $\nu=\nu_0\exp[-E_a/k_B(T_v-T_0)]$, does not yield a stable or physically meaningful Vogel–Fulcher temperature $T_0$, disfavoring an interpretation in terms of a cooperative glassy transition with a well-defined freezing temperature. 
Such slow dynamics can naturally coexist with the absence of a sharp thermodynamic anomaly and with the persistence of fast spin fluctuations, such as inelastic neutron scattering (INS), supporting a spin-liquid–proximate regime in which pronounced slowing down appears only at the longest time scales accessible to low-frequency susceptibility measurements.

This type of very low-temperature frequency dependent peak has been seen before in triangular QSL candidates \cite{Ma2018,Bag-Haravifard_2024,belbase2025field,kengle2026randomsinglet}, but it remains an open question how to interpret. Frequency-dependent susceptibility peaks are characteristic of spin glasses \cite{RevModPhys.58.801}, but the temperature $\frac{40 \> {\rm mK}}{J_1/k_B} \approx 1/160$ is much lower than the energy scale of the exchange, and the degree of correlations above $T_f$ is very strong. Thus the ground state likely includes a significant degree of quantum entanglement, perhaps in a two-dimensional version of a random-singlet phase \cite{kengle2026randomsinglet} driven by small randomness in the ground state. 
When the excitation field is lowered to 0.01~Oe (Fig. \ref{fig:susceptibility}\textbf{c}), the peak persists, indicating it is an intrinsic zero-field feature. Below 40~mK the susceptibility continues to decrease down to 0.7~mK with no clear features.   



The key question about NaYbSe$_2$ is whether it lies within the QSL phase. The data here are consistent with NaYbSe$_2$ being within the QSL phase above $T=40$~mK, given the complete lack of magnetic order. 
{A further piece of evidence in favor of QSL physics is that the quantum critical effects seen in KYbSe$_2$ are suppressed in NaYbSe$_2$. More specifically, the neutron spectra in KYbSe$_2$ show energy temperature scaling \cite{scheie2024_KYS} due to the proximity to the quantum critical point (QCP) between the 120$^{\circ}$ and QSL phase (see Figure \ref{fig:Schematic}). Quantum Fisher Information is a sensitive gauge to quantum criticality \cite{hauke2016,laurell2024witnessing} and the elevated value of $\rm nQFI = 3.4(2)$ \cite{scheie2024_KYS} in KYbSe$_2$ indicates the influence of the QCP. In contrast $\rm nQFI = 2.3(5)$ for NaYbSe$_2$ (see Supplementary Material)}
{is consistent with the material being beyond the QCP  (where nQFI should be a maximum) and within the QSL phase where spectral intensity is more distributed \cite{Sherman_2023_spectral,drescher_dynamical_2023}.}

However, the presence of a low-temperature frequency-dependent peak complicates our understanding. For triangular magnets that do not magnetically order, such a feature appears to be ubiquitous (c.f. YbMgGaO$_4$ \cite{Ma2018}, ${\mathrm{YbZn}}_{2}{\mathrm{GaO}}_{5}$ \cite{Bag-Haravifard_2024}, and TlYbSe$_2$ \cite{belbase2025field}, and YbCu$_x$Se$_2$ \cite{kengle2026randomsinglet}). 
This may indicate an inherent instability in the triangular QSL phase, as suggested by $\mathbb{U}_1$ Dirac simulations~\cite{pollmannvbs}. In this case, the weak singlet-forming distortions occur in a random pattern around dilute impurities. Then two possibilities emerge: glassy frozen regions nucleate around impurities (which are always present in real materials) or this small degree of randomness disrupts the entire phase. Resolving this question will require further experiments, but at a minimum the data here  evidence a coherent QSL for $T > 40$~mK---much lower temperature than the proximate QSL in KYbSe$_2$ \cite{scheie2024_KYS}.

In conclusion, we have used neutron spectroscopy, heat capacity, and magnetic susceptibility to investigate NaYbSe$_2$. 
We observe evidence for QSL physics in that (i) the coherent excitations observed in the neutron spectra are consistent with QSL simulated spectra, (ii) no static magnetic order is observed in specific heat down to 100~mK, and (iii) quantum entanglement witnesses indicates NaYbSe$_2$ has less divergent intensity than KYbSe$_2$, and is within the QSL phase. 
However, the frequency-dependent susceptibility peak indicates that an unexpected state forms below $T=40$~mK. This presumably is driven by some randomness in the exchanges on the order of 35~$\mu$eV. 
Above this energy scale however, the system acts like the long-sought triangular QSL phase. 


\section*{Acknowledgments}
 The work by A.O.S., M.L., K.W., S.T., V.S.Z., C.D.B., J.E.M., and D.A.T. is supported by the Quantum Science Center (QSC), a National Quantum Information Science Research Center of the U.S. Department of Energy (DOE). 
 The neutron scattering study used resources at the Spallation Neutron Source, a DOE Office of Science User Facility operated by the Oak Ridge National Laboratory under IPTS-30453 and IPTS-28649.  
  The work of P.L. and E.D. was supported by the U.S. Department of Energy, Office of Science, Basic Energy Sciences, Materials Sciences and Engineering Division. 
  The work by H.Z. is supported by the U.S. Department of Energy under Grant No. DE-SC0020254. 
A portion of this work was performed at the National High Magnetic Field Laboratory, which is supported by National Science Foundation Cooperative Agreement No. DMR-2128556 and the State of Florida. 
The work by Q.Z. and J. Ma was supported by the National Key Research and Development Program of China (Grant No. 2022YFA1402700). 
We acknowledge helpful discussions with  Johannes Knolle.


\newpage
\section*{Supplemental Materials for Spectrum and low-energy gap in triangular quantum spin liquid $\rm NaYbSe_2$}

\section{Sample Synthesis}

The samples for the neutron experiments were grown with NaCl flux and are the same as reported in Ref. \cite{Scheie_2024_Nonlinear}.
A new batch of samples were grown for the susceptibility measurement. A mixture of 1.58 gram NaCl powder, 0.23 gram Yb pieces, and 0.26 gram Se pieces were sealed in a vacuumed quartz tune. The tube was vertically located in a box furnace. The temperature profile for the reaction is that the temperature was raised to 850 Celsius degree with 50 degree/hour rate, stayed 16 days, and then decreased to 750 Celsius degree with 1 degree/hour rate, and thereafter decreased to room temperature with 100 degree/hour rate. The reddish thin plates of crystals could be picked out after the whole product was washed by water.

\section{Neutron Experiments}

We measured the inelastic spectrum of NaYbSe$_2$ using the $\sim 300$~mg co-aligned sample used in Ref.~\cite{Scheie_2024_Nonlinear} mounted in a dilution refrigerator (no magnet was used in this experiment).
We measured the $hhl$ inelastic scattering on the CNCS spectrometer~\cite{CNCS} at Oak Ridge National Laboratory's Spallation Neutron Source~\cite{mason2006spallation}, measuring at $E_i=3.32$~meV, $1.55$~meV, and 1.0~meV, rotating 180$^{\circ}$ to map the neutron spectrum. 
We measured at $T=0.1$~K and 12~K for a background. 
The data are shown in main text Fig. 2, and were normalized to absolute units by normalizing the magnon mode measured in Ref.~\cite{Scheie_2024_Nonlinear} to the nonlinear spin wave theory, such that the effective spin is 1/2 and is amenable to QFI calculation \cite{scheie2025tutorial}. 

\begin{figure*}
	\centering
	\includegraphics[width=0.8\textwidth]{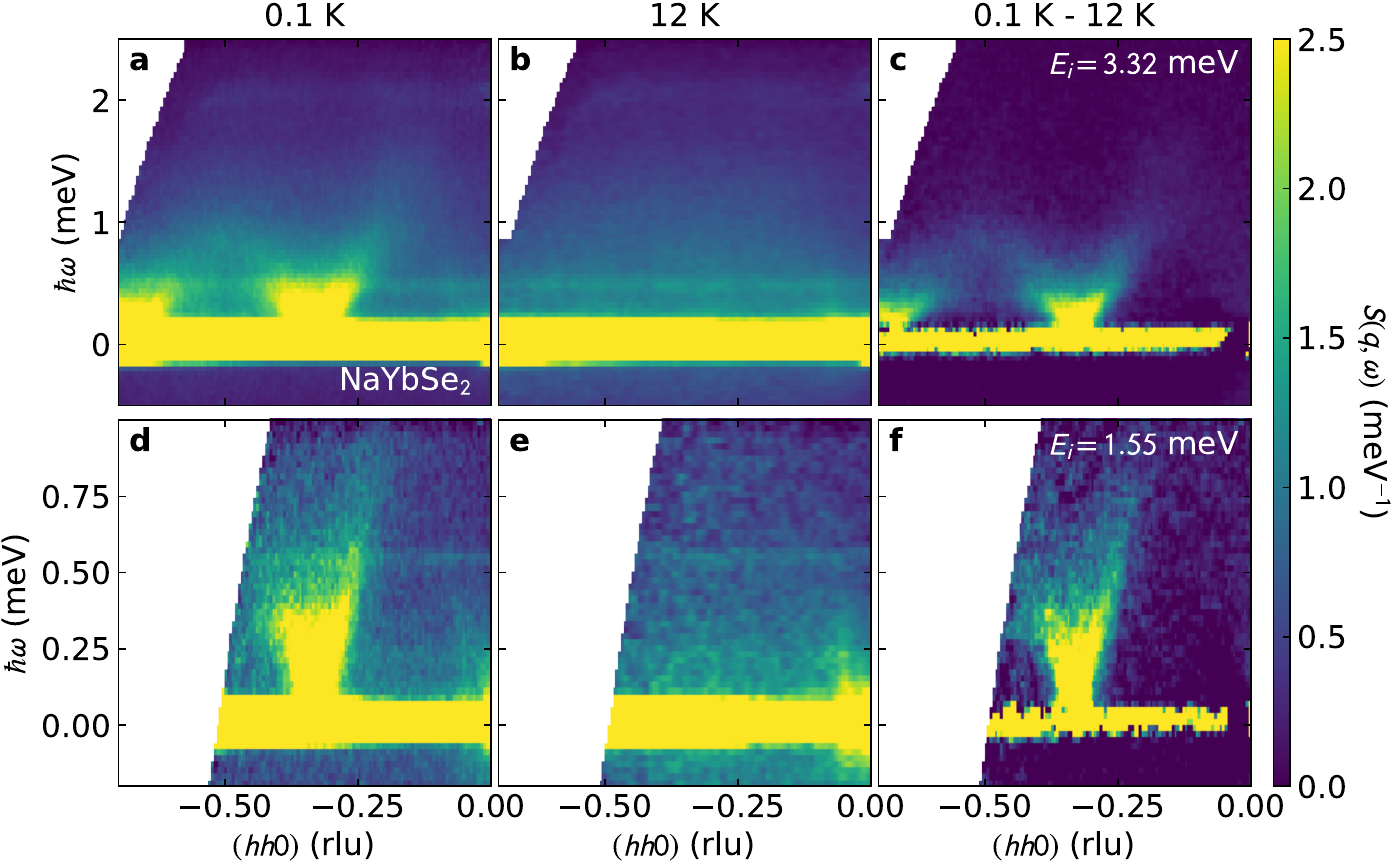}
	\caption{Neutron scattering data on NaYbSe$_2$ in the $hh\ell$ scattering plane, integrated over $\ell < 4.5$ reciprocal lattice units (rlu). The top row (a)-(c) shows scattering with $E_i = 3.32$~meV, the bottom row (d)-(f) shows scattering with $1.55$~meV. The left column shows the raw data at 0.1~K, the middle column the background at 12~K, the right column shows the background subtracted data.
    }
	\label{fig:NeutronSubtraction}
\end{figure*}

\begin{figure}
	\centering
	\includegraphics[width=0.5\textwidth]{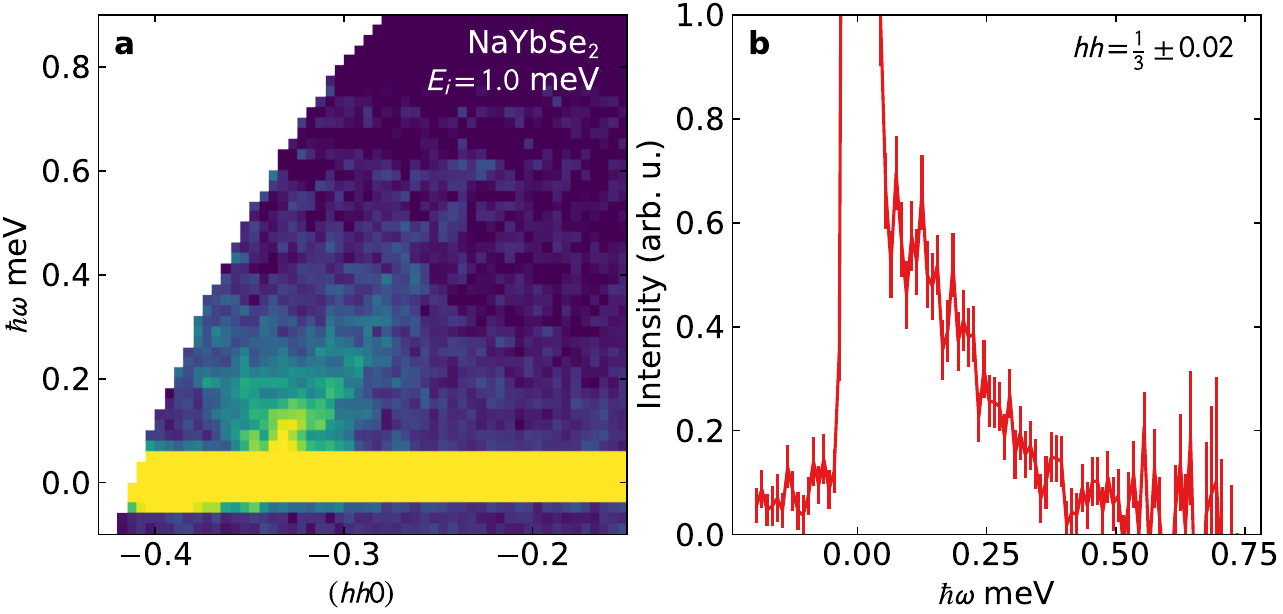}
	\caption{NaYbSe$_2$ scattering with $E_i = 1.0$~meV and $|\ell| \leq 1.0$ rlu. Panel (a) shows data along $hh$, panel (b) shows a constant $Q$ cut at $hh = (1/3,1/3)$. Intensity monotonically decreases with increasing energy, indicating a gapless spectrum to within $\pm 0.02$~meV.}
	\label{fig:Kcut}
\end{figure}

Figure \ref{fig:Kcut} shows the inelastic spectrum with an incident energy $E_i = 1.0$~meV, which gives an elastic line FWHM energy resolution 0.02~meV. With this resolution, no gap is observed at $K$.

\begin{figure*}
	\centering
	\includegraphics[width=0.6\textwidth]{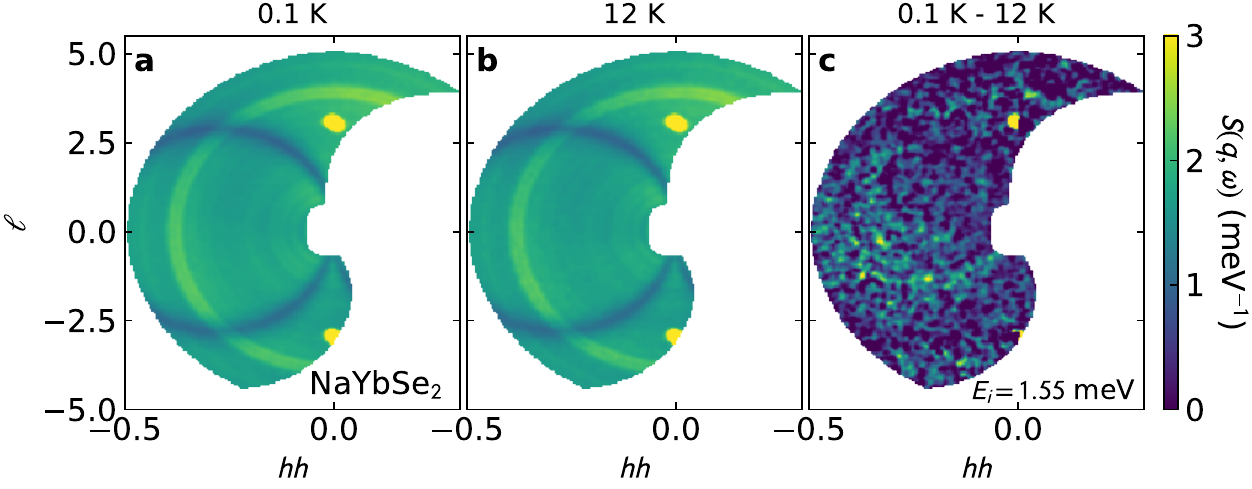}
	\caption{NaYbSe$_2$ elastic scattering with $E_i=1.55$~meV at 0.1~K (a), 12~K (b), and 0.1~K $-$ 12~K (c), with an energy window $\pm 0.04$~meV. No static spin correlations are visible in the temperature-subtracted data, suggesting an absence of long-range magnetic order. Note that in the unsubtracted data, there are arcs of suppressed intensity from when the vertical plates of the sample holder are along the incident and scattered beams respectively, and absorption is much larger.}
	\label{fig:Elastic}
\end{figure*}

Figure \ref{fig:Elastic} shows the elastic scattering with the higher resolution $E_i=1.55$~meV data. Temperature-subtraction shows no elastic scattering at $hh=(1/3,1/3)$, indicating an absence of long range static magnetic order. However, this may be because the CNCS spectrometer is not sensitive enough: similar CNCS scans on KYbSe$_2$ showed no static magnetism at zero-field \cite{Scheie_2024_Nonlinear}, even though triple axis scans clearly showed the onset of elastic Bragg intensity \cite{scheie2024_KYS}. Therefore, the absence of detectable NaYbSe$_2$ elastic scattering in these data does not necessarily indicate the absence of static magnetic order. 

\begin{figure*}
	\centering
	\includegraphics[width=0.93\textwidth]{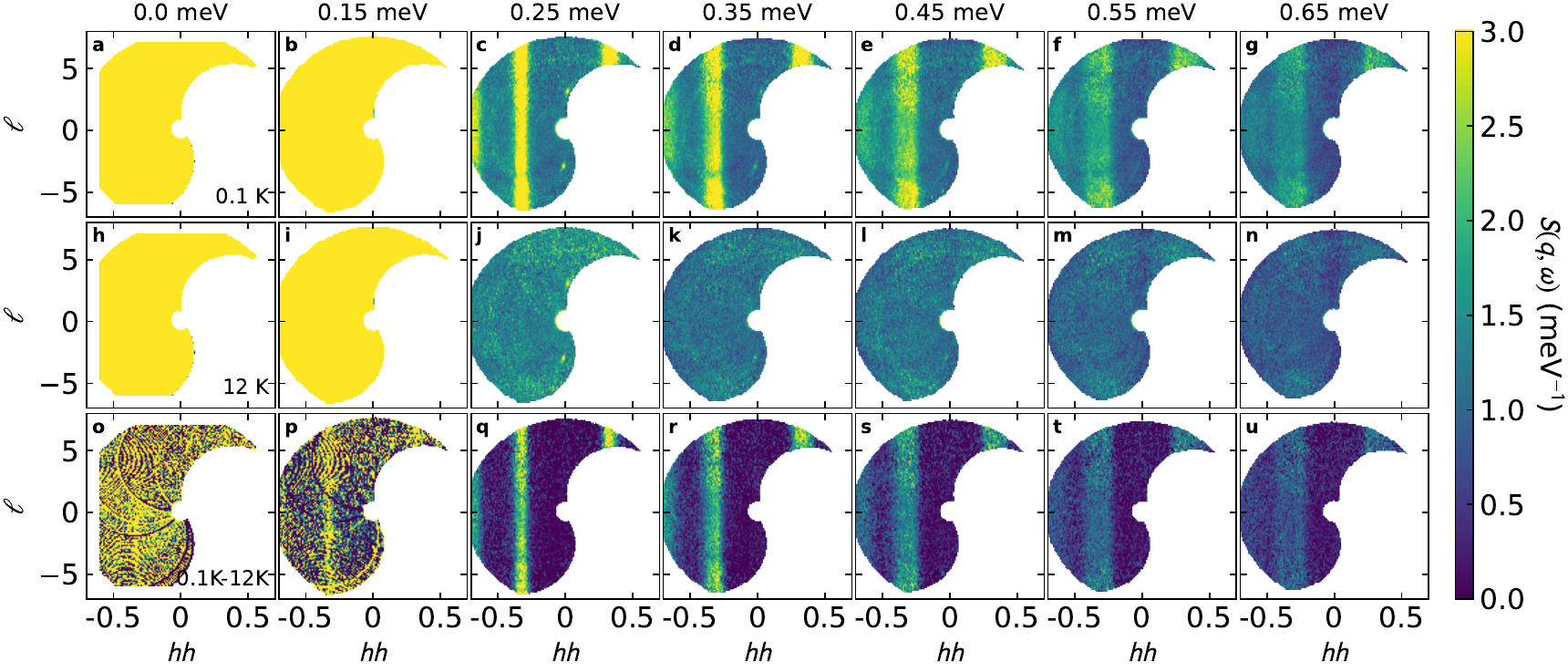}
	\caption{Constant energy slices of NaYbSe$_2$ with $E_i=3.32$~meV. The top row (a)-(g) shows the 0.1~K data, the middle row (h)-(n) shows the 12~K background, and the bottom row (o)-(u) shows the background subtracted data. Note the vertical streaks in the top and bottom row which reveals spin correlations independent of $\ell$, meaning the magnetic excitations are two-dimensional and have no correlations between triangular lattice planes.}
	\label{fig:ConstE}
\end{figure*}

\begin{figure*}
	\centering
	\includegraphics[width=0.8\textwidth]{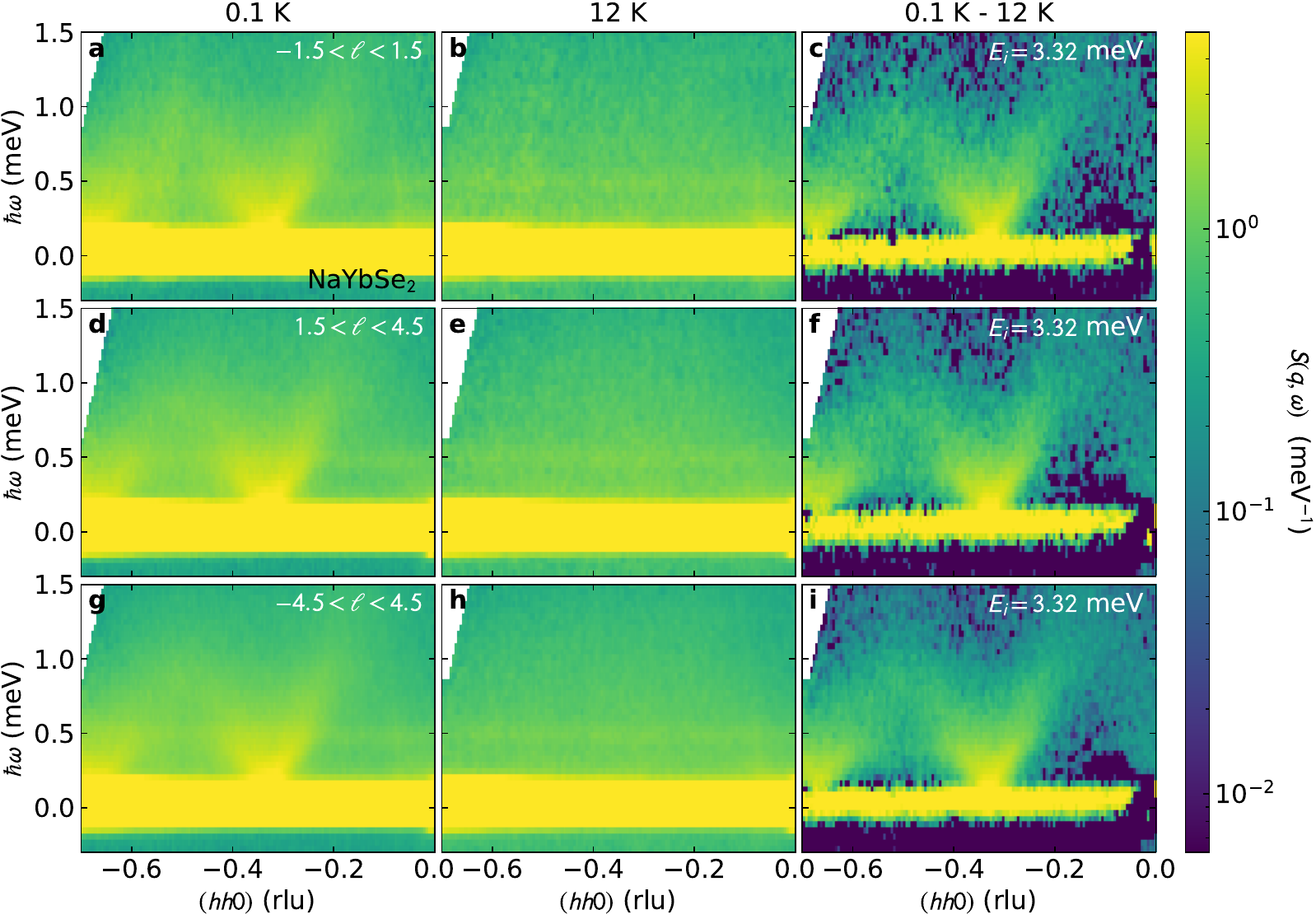}
	\caption{NaYbSe$_2$ inelastic scattering with different windows in $\ell$. The top row (a)-(c) shows $\ell$ centered around 0, the middle row (d)-(f) shows $\ell$ centered around 3, and the bottom row (g)-(i) shows $\ell$ integrated over both regions.}
	\label{fig:LdepSlices}
\end{figure*}

For a more complete view of the collected scattering data, Fig. \ref{fig:ConstE} shows constant energy slices of NaYbSe$_2$ with $E_i=3.32$~meV. Note the magnetic signal (most clearly shown in the temperature-subtracted data) has essentially no dependence on $\ell$, indicating no correlations between the triangular lattice planes.  Figure \ref{fig:LdepSlices} also shows this, with plots of different integration widths along $\ell$ which makes no visible difference to the inelastic scattering pattern. 
Therefore these scattering data are very two-dimensional. 

\begin{figure}
	\centering
	\includegraphics[width=0.45\textwidth]{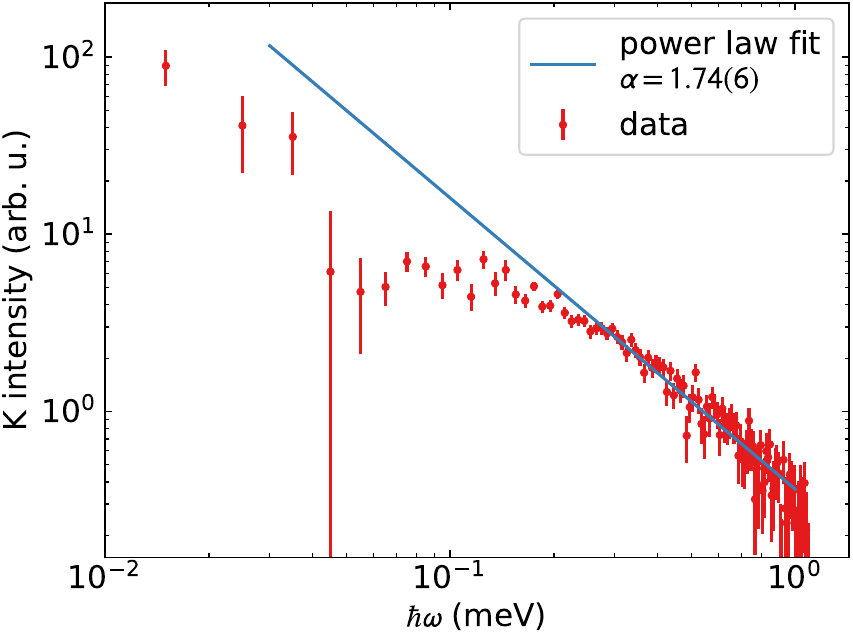}
	\caption{Temperature-subtracted NaYbSe$_2$ neutron scattering at $K$. Some power law behavior appears at high energy transfers, but it seems to deviate from this at low energy transfers. The fitted exponent strongly depends upon the region fitted, but the higher energy region follows $\alpha = 1.74(6)$.}
	\label{fig:criticalscaling}
\end{figure}

Figure \ref{fig:criticalscaling} shows the intensity at $K$ as a function of energy transfer. Unfortunately, because only one temperature is available, it is not possible to evaluate the presence or absence of a power-law scaling collapse to the data. Instead, we merely point out that the high energy transfer region appears to follow a power law with $\alpha = 1.74(6)$, consistent with the fitted KYbSe$_2$ value of $\alpha = 1.73(12)$ \cite{scheie2024_KYS} (though the precise exponent depends upon the fitted energy transfer region).

\section{AC Calorimetry}

Ac calorimetry measurements under hydrostatic pressure were performed in a piston-clamp pressure cell using Daphne oil 7373 as the pressure medium using the standard steady state technique \cite{PhysRev.173.679}. The temperature oscillations were measured using an Au/0.07\%Fe-chromel thermocouple, and a constantan meander was attached to the opposite side of the sample to apply heat. 
The heater power was varied between 25~nW and 5~$\rm \mu$W depending on the sample temperature. The measurement frequency was continuously adjusted to keep a constant phase relationship between the applied heat and the temperature oscillations on the thermocouple. For the lowest ($\leq$~100~nW) powers and temperatures the frequency was fixed near 2 Hz because the signal was too small to continuously vary the frequency. 
Below 300~mK, it was not possible to find a frequency range where $f\Delta{}T_{ac}$ was constant. This indicates that the internal relaxation of the sample is likely slower than the relaxation rate to the bath. Nonetheless, the measurement would still be sensitive to phase transitions even in this temperature range.

\begin{figure}
	\centering
	\includegraphics[width=0.45\textwidth]{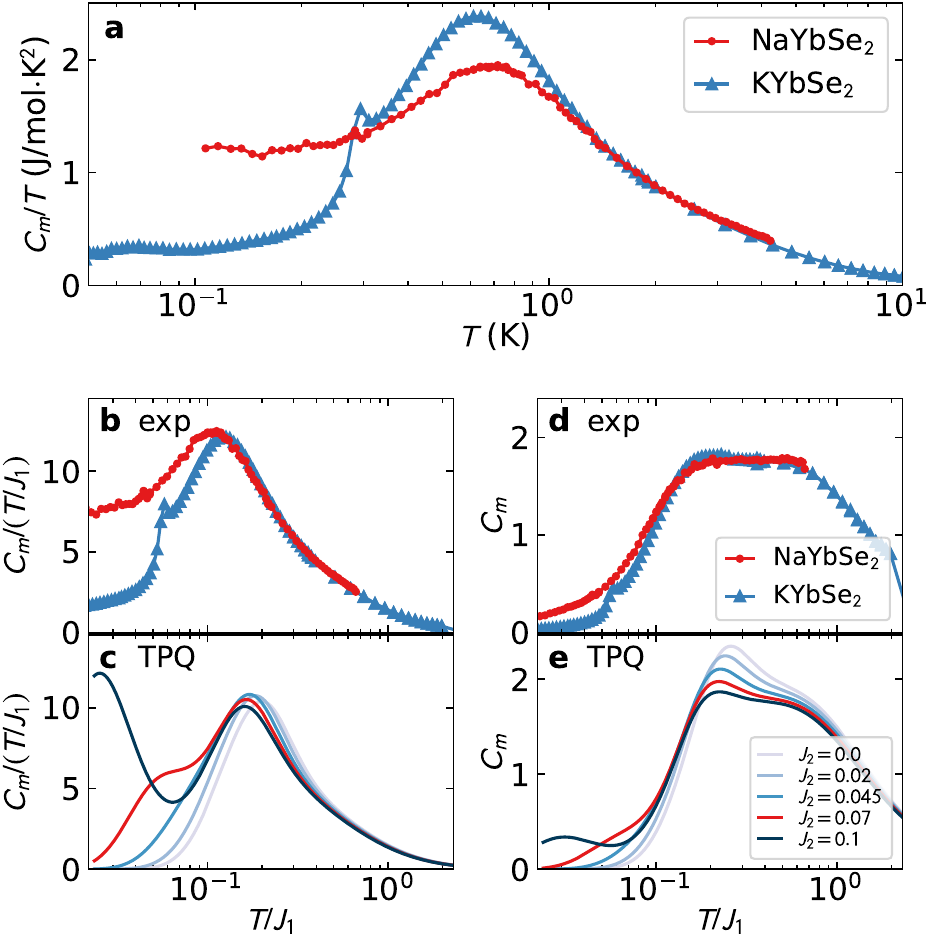}
	\caption{NaYbSe$_2$ specific heat compared to KYbSe$_2$  (Schottky anomaly subtracted) from Ref. \cite{scheie2024_KYS}. For NaYbSe$_2$ the nonmagnetic NaLuSe$_2$ specific heat \cite{PhysRevMaterials.3.114413} was subtracted. Panel \textbf{a} shows the specific heat, and panels \textbf{b} and \textbf{d} show the data with the temperature axis scaled by fitted $J_1$ \cite{Scheie_2024_Nonlinear}, plotted as $C$ and $C/T$ respectively. Panels \textbf{c} and \textbf{e} show the TPQ calculated specific heat as a function of $J_2$ (in units of $J_1$), with the value closest to the fitted NaYbSe$_2$ shown in red. }
	\label{fig:HCtheory}
\end{figure}

In Fig. \ref{fig:HCtheory} the NaYbSe$_2$ specific heat is compared to previously published KYbSe$_2$ data \cite{scheie2024_KYS}. The ``bump'' in $C/T$ is smaller in NaYbSe$_2$ than in KYbSe$_2$, while the specific heat below 300~mK is significantly larger in NaYbSe$_2$. This indicates more of the density of states has shifted to low energies, which is consistent with the system being within a QSL phase. 
Note also, in main text Fig. 3 there is no significant missing entropy in  NaYbSe$_2$, which again is consistent with it being in a well-defined quantum ground state rather than a non-ergodic glassy frozen state. 


Figure \ref{fig:HCtheory} also shows the experimental data from  NaYbSe$_2$ and KYbSe$_2$ compared to the TPQ simulations. In $C/T$ the theoretical heat capacity maximum is at higher temperature than the experimental maximum, possibly due to a finite-size-induced gap. However, on a qualitative level the resemblance between theory and experiment is strong, and the theoretical trend is consistent with NaYbSe$_2$ having a larger second neighbor exchange $J_2$ than KYbSe$_2$. 

\section{Ac susceptibility}
\subsection{Method}
The ac susceptometer comprises a solenoidal coil to generate an ac magnetic field and a pair of sensing coils housed within it. The pair of sensing coils are wound in opposite directions, ensuring they possess equal mutual inductance in magnitude but opposite signs. Consequently, when two sensing coils are connected in series, the induced voltage across them becomes zero. The presence of a sample positioned in the center of one of the sensing coils induces a nonzero net voltage across the coils. This induced voltage is directly proportional to the change in magnetic flux passing through the sensing coil over time. More detailed information can be found in https://nationalmaglab.org/user-facilities/dc-field/measurement-techniques/ac-magnetic-susceptibility-dc/. 
This setup includes a nonzero background susceptibility. Based on our experience of running this setup for over ten years, we believe that the excessive susceptibility near zero magnetic field is due to coil background, although we did not perform a background measurement. The background in temperature scans is much smaller compared to the sample signal. Therefore, the susceptibility peak at 40~mK is due to the sample's intrinsic behavior (confirmed by the absence of such a downturn in KYbSe$_2$ data, see below). We used ``Arbi. Unit'' because of the background signal of the AC susceptometer. 

To confirm the intrinsic nature of the 40~mK peak we also measured ultra-low temperature susceptibility down to 0.7~mK at the High B/T facility. This measurement used a much smaller excitation field of 0.01~Oe, but the peak is still present (albeit much less dramatic and at slightly higher temperature). Thus we associate the low temperature susceptibility downturn to intrinsic NaYbSe$_2$ behavior. 
More detailed information can be found in https://nationalmaglab.org/user-facilities/high-b-t/magnets-instruments/8-tesla-bay-2/

\subsection{Additional data}

\begin{figure}
	\centering
	\includegraphics[width=0.5\textwidth]{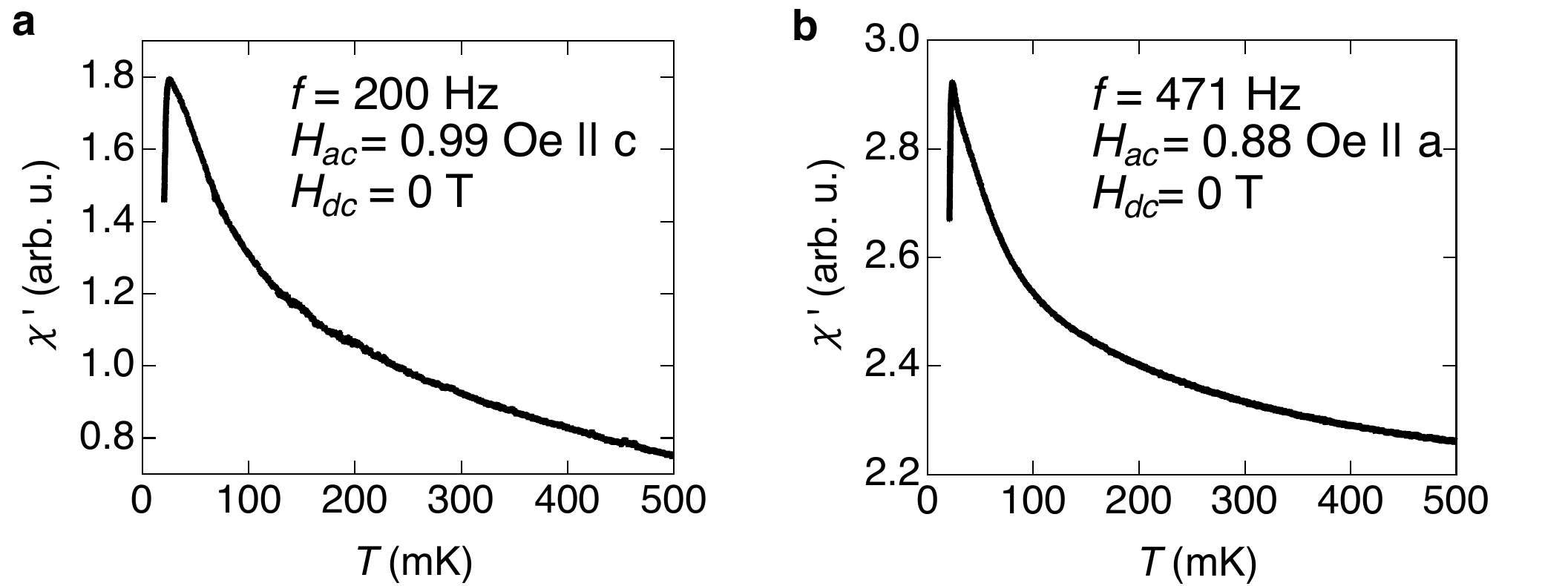}
	\caption{Temperature-dependent NaYbSe$_2$ ac susceptibilities at zero dc magnetic field with the ac field along (a) the {\it a}-axis and (b) the {\it c}-axis up to 500 mK. Above 25 mK, the susceptibility shows a gradual decrease with increasing temperature, indicative of paramagnetic behavior.}
	\label{fig:ac_zerofield_highT}
\end{figure}

Figure \ref{fig:ac_zerofield_highT} shows the temperature-dependent AC susceptibility at zero magnetic field up to higher temperatures than in the main text Fig. 4. Paramagnetic behavior is evident up to 500~mK, with no phase transitions visible. 
Note that with this higher excitation field, there appears to be sample heating which shifts the susceptibility downturn lower in temperature and makes it asymmetric (which we initially misinterpreted as a gap).

\begin{figure}
	\centering
	\includegraphics[width=0.5\textwidth]{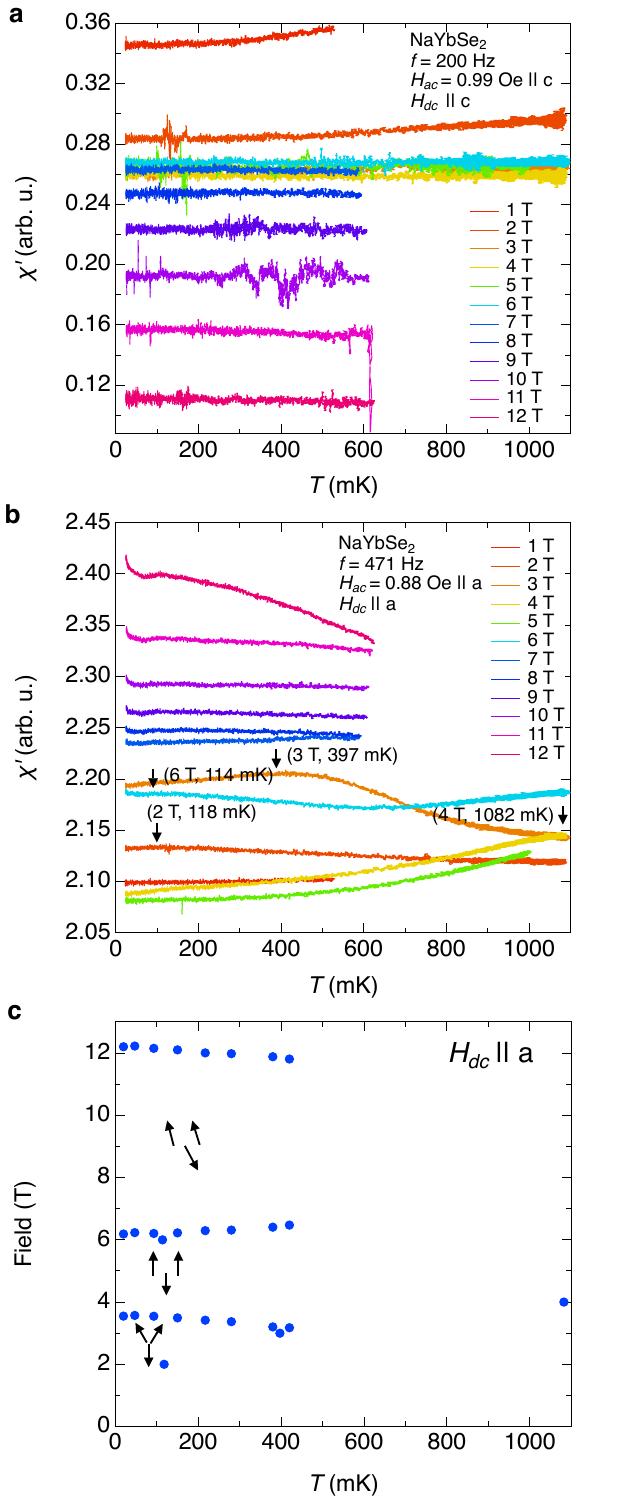}
	\caption{(a) AC susceptibility of NaYbSe$_2$ vs. temperature at fields parallel to the $c$-axis from 1 T to 12 T. (b) AC susceptibility vs. temperature at fields parallel to the $a$-axis from 1 T to 12 T. The arrows indicate the phase transition points with respect to dc field and temperature. (c) A magnetic phase diagram of NaYbSe$_{2}$ field-dependent transitions with the field along the $a$-axis. The data points were taken from AC susceptibility measurements. The arrows represent a schematic of the spin structure for each phase.}
	\label{fig:ac_HF_PD}
\end{figure}

Figure \ref{fig:ac_HF_PD} shows additional temperature-dependent NaYbSe$_2$ susceptibility data for applied fields between 1~T and 12~T. For field applied along $c$, there are no clear features in the data indicating phase boundaries. For field along $a$, there are several kinks and discontinuities. The phase diagram from temperature and field dependent susceptibility features is plotted in panel (c) of Fig. \ref{fig:ac_HF_PD}. 

\begin{figure}
	\centering
	\includegraphics[width=0.5\textwidth]{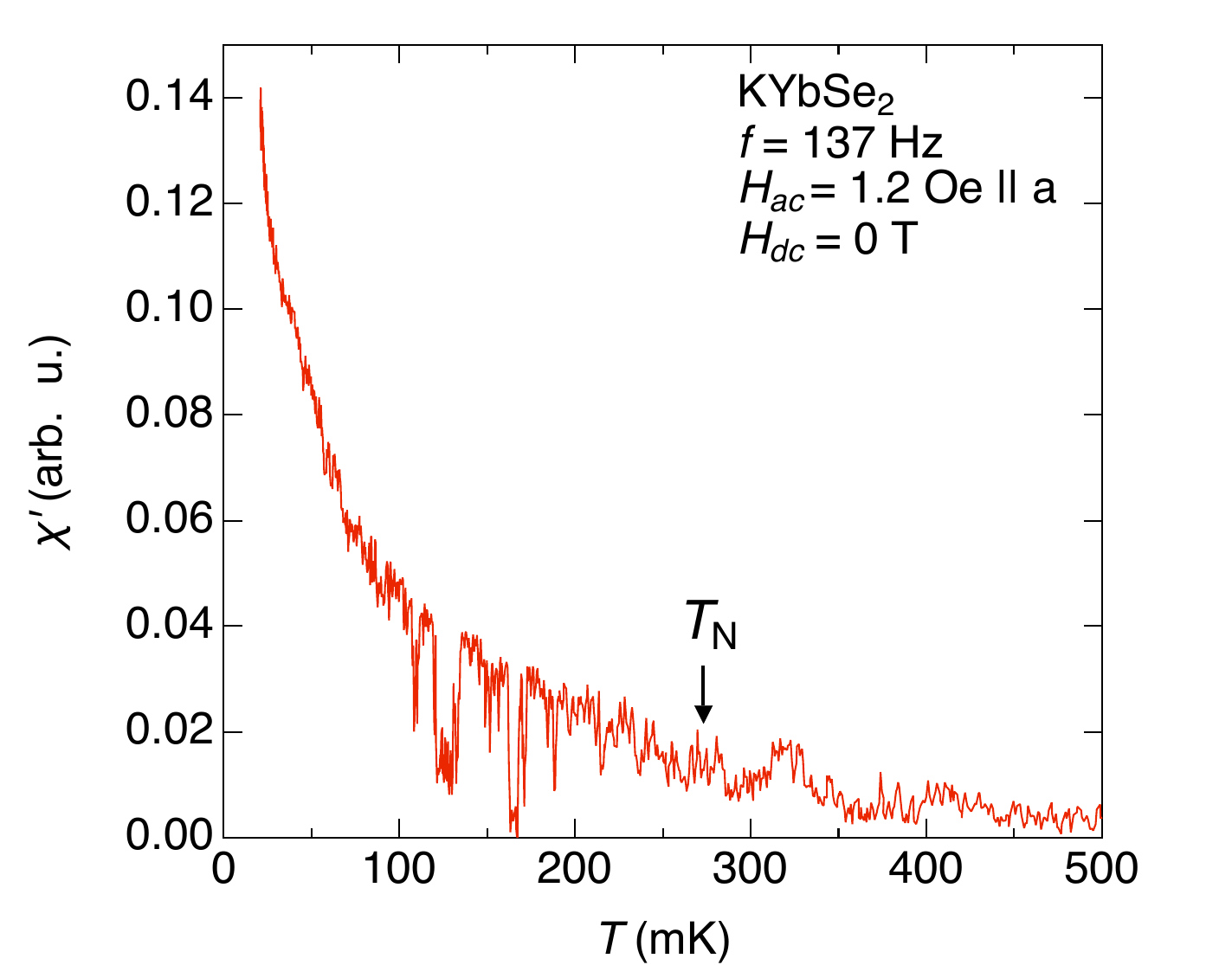}
	\caption{AC susceptibility of KYbSe$_{2}$. The feature near the ordering temperature $T_{\text{N}} \sim 290$ mK is negligible in AC susceptibility because the magnetic moment is suppressed by strong quantum fluctuations.
}
	\label{fig:KYS_susceptibility}
\end{figure}

Finally, for comparison with NaYbSe$_2$, Figure \ref{fig:KYS_susceptibility} shows the measured in-plane susceptibility of KYbSe$_{2}$ (which was also measured in the same cryostat at the same time---and therefore the same temperature and field configurations---as the two NaYbSe$_2$ crystals). Note the absence of a downturn in the data, which follows a $1/T$ divergence to the lowest temperatures. Note also that the ordering transition is not visible in the data (which is admittedly somewhat noisy), consistent with the very subtle weak static magnetic order in this compound. 

\section{Theoretical simulations}

\subsection{MPS calculations}

We performed MPS simulations on the $J_2/J_1$ model with varying values of XXZ anisotropy $\Delta$ \cite{schollwock_density-matrix_2011,Sherman_2023_spectral,PhysRevB.107.L140411,drescher_dynamical_2023}.
$$H=J_1\sum_{\langle i,j\rangle}(S_i^xS_j^x+S_i^yS_j^y+\Delta S_i^zS_j^z)+J_2\sum_{\langle\langle i,j\rangle\rangle}(S_i^xS_j^x+S_i^yS_j^y+\Delta S_i^zS_j^z)$$
Simulations are done on a cylinder geometry with circumference $C=6$ and length $L=36$ with XC boundary conditions \cite{szasz_chiral_2020} on the triangular lattice, at a maximum bond dimension of $\chi=512$ using the \textit{ITensor} library \cite{fishman_itensor_2022}. The ground state $|\Omega\rangle$ of the model is found using the density matrix renormalization group (DMRG). The spin-spin correlation function is determined with time evolution using the time-dependent variational principle (TDVP) with a time step of $dt=0.1$ \cite{haegeman_time-dependent_2011,haegeman_post-matrix_2013,haegeman_unifying_2016,vanderstraeten_tangent-space_2019,yang_time-dependent_2020,Sherman_2023_spectral}.
$$G(\mathbf{x},t)=\langle \Omega|\mathbf{S}_\mathbf{x}(t)\cdot \mathbf{S}_c(0)|\Omega\rangle$$
where the subscript $c$ represents the central site on the cylinder. The dynamical spin spectral function is then computed as the Fourier transform of the correlation function.
$$S(\mathbf{x},t)=\frac{1}{N}\sum_\mathbf{x}\int_0^\infty \frac{dt}{2\pi}e^{i(\mathbf{q}\cdot\mathbf{x}-\omega t)}G(\mathbf{x},t)$$
To remedy the finite time cutoff of the Fourier transform, Gaussian broadening of the time data---on the order of the cutoff $T_{\text{max}}\sim 80$---is applied to the correlation function before transforming \cite{Sherman_2023_spectral}.

\subsection{TPQ specific heat calculations}

\begin{figure}
	\includegraphics[width=\columnwidth]{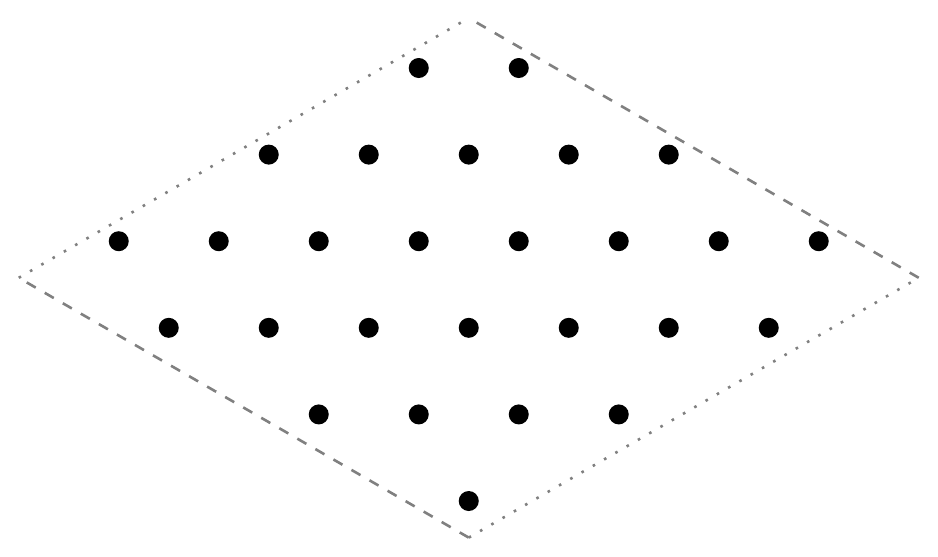}
	\caption{Finite size cluster used for the TPQ calculations. The number of sites (27) is divisible by three to be compatible with the 120$^{\circ}$ order at low $J_2/J_1$. Periodic boundary conditions are applied across edges with dashed or dotted lines. \label{fig:SpinCluster}}
\end{figure}

We numerically calculated the magnetic specific heat $C_m$ for the $S=1/2$ AFM $J_1$-$J_2$ Hamiltonian
\begin{align}
	H	&=	J_1 \sum_{\langle i,j\rangle} \mathbf{S}_i \cdot\mathbf{S}_j  + J_2 \sum_{\langle\langle i,j\rangle\rangle}\mathbf{S}_i \cdot \mathbf{S}_j
\end{align}
on a 27-site cluster (shown in Fig. \ref{fig:SpinCluster}) with periodic boundary conditions  using the microcanonical thermal pure quantum state (TPQ) \cite{PhysRevLett.108.240401} method and the  $\mathcal{H}\Phi$ library \cite{Kawamura2017, Ido2024}, version 3.5.2. In this typicality-based approach, a thermal quantum state is iteratively constructed starting from a randomized initial vector, and associated with a temperature estimated from the internal energy. To reduce statistical errors, we averaged over 15 initial vectors. Finite-size errors are expected to mainly affect the results at low temperatures \cite{PhysRevB.98.035107, PhysRevResearch.2.013186}, but not to change the trend with $J_2/J_1$ highlighted here.

\subsection{Phase transition through neural quantum states (NQSs)}

\subsubsection{NQS wave function}
The NQS method utilizes an artificial neural network as a variational wave function to approximate the ground state of a target model~\cite{Carleo_Science17_NQS}. 
In a system with $N$ spin-1/2 degrees of freedom, the Hilbert space can be spanned by the $S_z$ basis 
$\ket{\sigma} = \ket{\sigma_1,...,\sigma_N}$ with 
$\sigma_i = \,\uparrow$ or $\downarrow$. 
Similar to image recognition tasks in which the artificial neural network converts every image input to a probability, in quantum many-body problems the NQS converts every input basis $\ket{\sigma}$ to a wave function amplitude $\psi_\sigma$. This gives the full quantum state as
\begin{equation}
    \ket{\Psi} = \sum_\sigma \psi_{\sigma} \ket{\sigma}.
\end{equation}

\begin{figure}[b]
    \centering
    \includegraphics[width=0.95\linewidth]{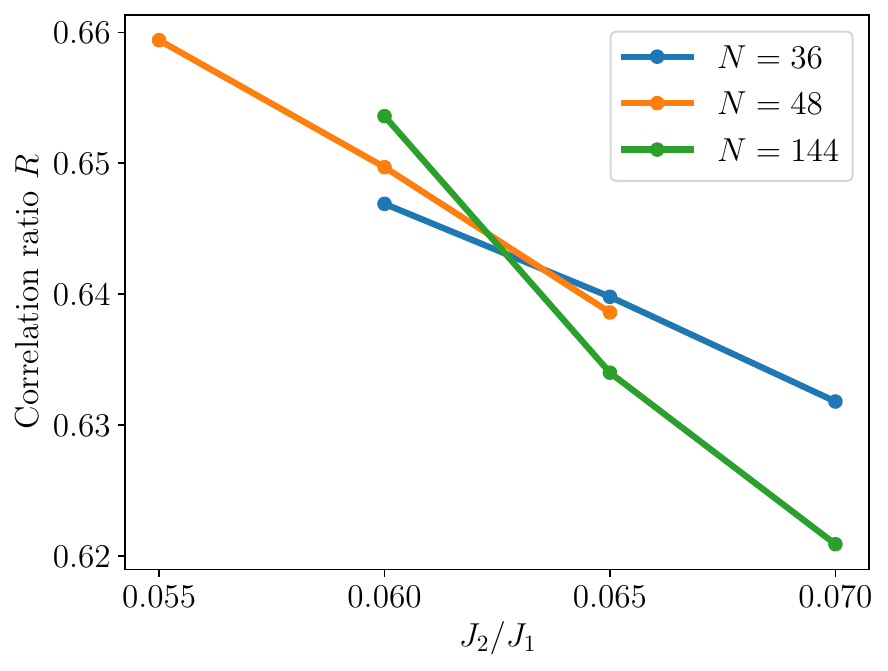}
    \caption{Correlation ratio}
    \label{fig:correlation_ratio}
\end{figure}

In this work, we employ deep residual convolutional neural networks as the variational wave function. The network contains 16 convolutional layers, each with 32 channels and $3\times3$ kernels, leading to 139008 real parameters in total. The GeLU activation is applied before each convolutional layer. The circular padding is utilized in the convolutional layer to realize the exact translation symmetry. The output after the last convolutional layer contains 32 channels, which is divided into two groups $x^{(1)}_j$ and $x^{(2)}_j$ each with 16 channels, and the final wave function amplitude output of the network is given by $\psi_\sigma = \sum_j \exp (x^{(1)}_j + i x^{(2)}_j)$, where we sum over all elements in the 16 channels. 

In addition, we apply symmetries on top of the well-trained $\psi_\sigma$ to project variational states onto suitable symmetry sectors. Assuming the system permits a symmetry group represented by operators ${T_i}$ with characters ${\omega_i}$, the symmetrized wave function is then defined as~\cite{Nomura_JPCM21_RBMsymm, Reh_PRB23_NQSsymm}
\begin{equation} \label{eq:symmetry}
    \psi^\mathrm{symm}_\sigma = \sum_i \omega_i^{-1} \psi_{T_i \sigma}.
\end{equation}
The applied symmetry groups in Eq.\,\eqref{eq:symmetry} are the $D_6$ group realizing rotation and reflection symmetries and the $Z_2$ group realizing the spin inversion symmetry $\sigma \rightarrow -\sigma$.

The deep network is trained by the MinSR method to approach the ground state of the triangular $J_1$-$J_2$ Hamiltonian~\cite{Chen_arxiv23_MinSR}. The training employs $10000$ Monte Carlo samples, 20000 steps without symmetries followed by 10000 steps with symmetries.

\subsubsection{Phase transition}
The transition between the $120^\circ$-ordered and the QSL phase can be detected through the spin structure factor
\begin{equation}
    S(\mathbf{q}) = \frac{1}{N} \sum_{ij} C_{ij} e^{i \mathbf{q} \cdot (\mathbf{r}_i - \mathbf{r}_j)},
\end{equation}
where $\mathbf{q}$ denotes the momentum, and $C_{ij}$ is the real-space spin-spin correlation given by
\begin{equation}
    C_{ij} = \braket{\mathbf{S}_i \cdot \mathbf{S}_j},
\end{equation}
which is obtained from the NQS wave function by Monte Carlo sampling. The $120^\circ$ order is signaled by a peak in the spin structure factor $S(\mathbf{K})$ at $\mathbf{K} = (4\pi/3, 0)$. In the thermodynamic limit, $S(\mathbf{K})$ diverges only in the $120^\circ$ ordered phase but not in the QSL phase.

Importantly, the numerical simulations are performed for large but finite systems, leading to finite structure factors in both phases. In order to minimize finite-size effects for the detection of phase transitions, the so-called correlation ratio $R$ has been introduced~\cite{Kaul_PRL15_TriQSL, Pujari_PRL16_QBT, Nomura_PRX21_SquareQSL}
\begin{equation}
    R = 1 - \frac{S(\mathbf{K + \delta q})}{S(\mathbf{K})},
\end{equation}
where $\mathbf{K + \delta q}$ represents the nearest neighboring momentum of $\mathbf{K}$. The correlation ratio represents a measure for the sharpness of the spin structure factor. As the system size $N$ increases, $R$ grows in the $120^\circ$ ordered phase and decreases in the QSL phase. Most important for the current purpose, this opposite behavior in the two phases with system sizes, generically leads to a crossing point in $R$ for different $N$ at the phase transition point. As shown in Fig.\,\ref{fig:correlation_ratio}, the correlation ratio $R$ for different system sizes indeed exhibits such a crossing at $J_2/J_1 \approx 0.063$ signaling the phase transition.

We identify two sources for uncertainties in estimating the precise quantum
phase transition point, namely a variational bias and a statistical error.
First, for complex quantum models such as the considered frustrated magnets we
find that the variationally obtained wave function exhibits larger variational
errors upon increasing system size. We observe that these errors usually have
the tendency to lead to a stronger spin order and consequently to a larger
correlation ratio $R$ consistent with other works \cite{viteritti2024transformer}.
Therefore, our estimate for the phase transition point $J_2/J_1=0.063$ exhibits a
bias towards larger values of $J_2/J_1$ so that we interpret 0.063 as an upper bound.
Second, the measurement of $R$ is based on an underlying Monte Carlo sampling
scheme, which introduces statistical errors and leads to an uncertainty 0.001
in the critical $J_2/J_1$ value. In summary, the result provided in Fig. \ref{fig:correlation_ratio} lead
to a bound of the critical point of the form $J_2/J_1 \lesssim 0.063 \pm 0.001$.


%

\end{document}